\documentclass[10pt]{iopart}

\usepackage{iopams}  
\usepackage{pdfpages}
\usepackage{amstext} 
\usepackage{graphicx}                       
\usepackage{overpic}
\usepackage{array}  
\usepackage{hyperref} 
\hypersetup{colorlinks   = true,linkcolor    = blue}

\newcolumntype{C}[1]{>{\centering\let\newline\\\arraybackslash\hspace{0pt}}m{#1}} 
\newcolumntype{\empty}{@{}m{0pt}@{}} 

\begin{document}

\title{Fast physics-based launcher optimization for electron cyclotron current drive}
\author{N A Lopez$^1$, A Alieva$^1$, S A M McNamara$^1$ and X Zhang$^1$}
\address{$^1$ Tokamak Energy Ltd, Abingdon, United Kingdom}

\ead{nicolas.lopez@tokamakenergy.com}

\begin{abstract}

With the increased urgency to design fusion pilot plants, fast optimization of electron cyclotron current drive (ECCD) launchers is paramount. Traditionally, this is done by coarsely sampling the 4-D parameter space of possible launch conditions consisting of (1) the launch location (constrained to lie along the reactor vessel), (2) the launch frequency, (3) the toroidal launch angle, and (4) the poloidal launch angle. For each initial condition, a ray-tracing simulation is performed to evaluate the ECCD efficiency. Unfortunately, this approach often requires a large number of simulations (sometimes millions in extreme cases) to build up a dataset that adequately covers the plasma volume, which must then be repeated every time the design point changes. Here we adopt a different approach. Rather than launching rays from the plasma periphery and hoping for the best, we instead directly reconstruct the optimal ray for driving current at a given flux surface using a reduced physics model coupled with a commercial ray-tracing code. Repeating this throughout the plasma volume requires only hundreds of simulations, constituting a significant speedup. The new method is validated on two separate example tokamak profiles, and is shown to reliably drive localized current at the specified flux surface with the same optimal efficiency as obtained from the traditional approach.

\end{abstract}

%
%
%
%
\ioptwocol

\section{Introduction}

Tokamak Energy has recently begun the process of designing a fusion pilot plant (FPP) based on the spherical tokamak (ST) concept~\cite{Peng86,Stambaugh98} as part of the U.~S. Department of Energy's Milestone-Based Fusion Development Program~\cite{Hsu23,Kingham24}. Due to their lower aspect ratio, STs have reduced space for a central solenoid; this limitation is exacerbated when one also accounts for the additional inboard shielding necessary to protect high-temperature superconducting magnet coils from neutron damage~\cite{Menard19}. Therefore, non-solenoidal means of driving current are viewed as necessary features of ST-based FPPs. Electron cyclotron current drive (ECCD)~\cite{Erckmann94, Prater04} is a promising option for this task because, with proper launching parameters, it can drive localized current throughout the plasma volume to enable current profile tailoring for advanced operations~\cite{Citrin10}. Additionally, the ECCD launchers take up only a small fraction of the reactor blanket and are not as susceptible to neutron damage as other auxiliary current drive techniques.
 
Traditionally, the initial `zeroth-order' optimization of ECCD systems are performed using ray-tracing simulations in which the parameters governing the launcher (frequency, location, and launch angles) are varied in a coarse parameter scan to determine the optimal EC launcher design under some figure of merit. Examples of this traditional method of EC optimization abound, including on ST40~\cite{DuToit22}, NSTX-U~\cite{Lopez18a}, ITER~\cite{Farina14}, DEMO~\cite{EPoli13}, and STEP~\cite{Freethy24}. However, this optimization method is known for requiring a large number of simulations, sometimes millions in extreme cases, which is undesirably time-consuming. 

Here we demonstrate a faster, physics-based, alternative means of optimizing ECCD launchers. Our new method relies on the HARE reduced model~\cite{EPoli18b} for global ECCD estimation that has been successfully benchmarked on ITER and DEMO. In essence, we use the HARE model to first determine the necessary wave-particle interaction that yields optimal ECCD efficiency at a specified flux surface; we then infer the launcher configuration that would generate such a wave-particle interaction by coupling to a commercial ray-tracing code. The result is an optimization framework that uses, in principle, only four ray-tracing simulations per radial deposition location; adequately covering the plasma volume therefore requires on the order of hundreds, not millions, of simulations. As a result of this ten-thousand-fold speedup, we expect our new approach to be useful to the general fusion community in the quest to design FPPs.

This paper is organized as follows. In Sec.~\ref{sec:background}, the HARE reduced model is briefly reviewed, and its implementation into a fast physics-based ECCD launcher optimization scheme is outlined. In Sec.~\ref{sec:main}, the newly developed optimization scheme is compared to the traditional method of ECCD launcher optimization on a series of two example reactor-relevant plasma profiles. Finally, in Sec.~\ref{sec:concl} the main results are summarized.

\section{Theoretical background and optimization framework description}
\label{sec:background}

In this section we describe the underlying theoretical and modelling framework for our physics-based ECCD optimization scheme. We specifically consider only fundamental O-mode heating incident from the low-field side (LFS) of the tokamak periphery, as this will be the optimal scheme for our chosen example plasmas presented in the following section. In principle, higher-harmonic and/or X-mode heating schemes can be accommodated via straightforward modification of the absorption formula~\ref{eq:resE} (as described in Ref.~\cite{EPoli18b}), but we shall defer such investigations to future work.

\subsection{Summary of HARE reduced model for ECCD optimization}
\label{sec:HARE}

Here we first provide a high-level summary of the HARE reduced model for ECCD optimization, as it is needed for our work. For more details and discussion of the model limitations, the reader is invited to consult the original reference~\cite{EPoli18b}.

The wave-particle resonance condition for fundamental ECRH is given by
\begin{equation}
    \frac{\omega_c}{\omega} = \gamma - N_\parallel u_\parallel
    ,
    \label{eq:resCOND}
\end{equation}

\noindent where $\omega_c$ is the electron cyclotron resonance frequency and $u_{\perp,\parallel} = \gamma v_{\perp,\parallel} / c$ is the normalized relativistic velocity, with $\gamma$ being the relativistic Lorentz factor defined as
\begin{equation}
    \gamma =
    \frac{1}{\sqrt{1 - v_\parallel^2/c^2 - v_\perp^2/c^2}}
    \equiv
    \sqrt{1 + u_\parallel^2 + u_\perp^2}
    .
\end{equation}

\noindent For EC waves with $|N_\parallel| < 1$, the resonance curve in velocity space is an ellipse centered at $u_\perp = 0$ and $u_\parallel \neq 0$, with $\text{sign}(u_\parallel) = \text{sign}(N_\parallel)$. Hence, the electron distribution function (and thus the EC damping too) is expected to be maximized along the resonance curve at $u_\perp = 0$ where $|u_\parallel|$ attains its minimum value. One should note, however, that the resonance condition can only be satisfied if the magnetic field exceeds a threshold value given by the condition
\begin{equation}
    \frac{\omega_c}{\omega} \ge \sqrt{1 - N_\parallel^2}
    .
    \label{eq:onset}
\end{equation}

\noindent This condition is particularly important for EC waves incident from the LFS of the cyclotron harmonic as it determines the onset of EC damping.

Using these observation, one can derive estimates for the wave parameters $\omega$ and $N_\parallel$ that should yield efficient ECCD as follows. First, one computes the energy of the resonant electrons at peak damping as~\cite{EPoli18b}
\begin{eqnarray}
    E_{res}
    = T_e
    \log
    \left[
        \frac{\omega_p^2 \Delta}{\omega_c c}
        \sqrt{ \frac{T_e}{2\pi m c^2}}
    \right]
    ,
    \label{eq:resE}
\end{eqnarray}

\noindent where $\omega_p$ is the plasma frequency, $m c^2$ is the electron rest mass, and $\Delta$ is a free parameter governing the absorption width. In the original reference~\cite{EPoli18b}, $\Delta$ was set to the value $0.2 a$ for DEMO-like profiles, where $a$ is the minor radius. Here we reduce this to $\Delta = 0.1 a$ based on the heuristic estimate of Doppler-broadened ECRH, viz., $\Delta \propto R_{geo}$ with $R_{geo}$ the major radius of the plasma (engineering) geometric center; the value of $R_{geo} = 4.25~\text{m}$ we shall consider here is about half that of DEMO, while the values of $a$ are comparable. Note that Eq.~\ref{eq:resE} is based on the expression for fundamental O-mode heating provided in Ref.~\cite{Bornatici83}.

Next, having obtained $E_{res}$ via Eq.~\ref{eq:resE}, one uses the relation $E_{res} = mc^2(\gamma - 1)$ to obtain the corresponding resonant velocity for electrons with $u_\perp = 0$:
\begin{equation}
    |u_\parallel| = 
    \sqrt{
        \gamma^2 - 1
    }
    , \quad \gamma = \frac{E_{res}}{m c^2} + 1
    .
    \label{eq:resU}
\end{equation}

\noindent At this point there is a choice. For the simplified version of HARE presented in Ref.~\cite{EPoli18b}, one chooses a fixed value of $N_\parallel$. Then, $\omega$ is determined by imposing the resonance condition~\ref{eq:resCOND} at $u_\perp = 0$:
\begin{equation}
    \omega = \frac{
        \omega_c
    }{
        \gamma - N_\parallel u_\parallel
    }
    ,
    \label{eq:HAREsimpleOMEGA}
\end{equation}

\noindent where $u_\parallel$ and $\gamma$ are given in Eq.~\ref{eq:resU}. Intuitively, this formulation of HARE simply states that the optimal wave frequency for damping on a given target population of electrons [Eq.~\ref{eq:resU}] with given $N_\parallel$ is set by the relativistic Doppler shift with respect to the local cyclotron frequency. 

Alternatively, the complete HARE model presented in Ref.~\cite{EPoli18b} determines both $\omega$ and $N_\parallel$ by requiring that the peak absorption location (denoted $R_{dep}$) occurs near the ECRH onset location (denoted $R_{onset}$), as set by Eq.~\ref{eq:onset}. Specifically, one imposes that the distance between $R_{dep}$ and $R_{onset}$ is set by $\Delta$ as
\begin{equation}
    \mathcal{R}
    \doteq
    \frac{R_{dep}}{R_{onset}}
    \approx
    \left(
        1
        + \frac{\Delta}{R_{dep}} \cos \theta
    \right)^{-1}
    ,
\end{equation}

\noindent where $\theta$ is the inclination angle of the EC beam across the resonance (i.e. $\theta = 0$ for horizontal propagation). Assuming $\omega_c \sim 1/R$, one can relate $\mathcal{R}$ with $\omega$ and $N_\parallel$ via Eq.~\ref{eq:onset}, then subsequently isolate $N_\parallel$ to obtain
\begin{equation}
    N_\parallel = 
    \text{sign}(u_\parallel)
    \sqrt{
        1 
        - \mathcal{R}^2 \left( \frac{\omega_c}{\omega} \right)^2
    }
    ,
    \label{eq:NparFULLbad}
\end{equation}

\noindent where $\omega_c$ now corresponds to the cyclotron frequency at the peak deposition location. Inserting this expression for $N_\parallel$ into the resonance condition~\ref{eq:resCOND} then gives a quadratic equation for $\omega$, which can be solved to yield the relevant root~\cite{EPoli18b}
\begin{equation}
    \frac{\omega}{\omega_c} =
    \gamma
    + |u_\parallel| \sqrt{1 - \mathcal{R}^2}
    .
    \label{eq:HAREfullOMEGA}
\end{equation}

\noindent Lastly, having obtained $\omega/\omega_c$, one then can obtain $N_\parallel$ either via Eq.~\ref{eq:NparFULLbad} or through the simpler expression obtained from the resonance condition
\begin{equation}
    N_\parallel
    = \frac{
        \gamma
        - \omega_c/\omega
    }{
        u_\parallel
    }
    .
    \label{eq:HAREfullNpar}
\end{equation}

\noindent In summary, the complete HARE reduced model for ECCD sets $N_\parallel$ and $\omega$ by the simultaneous condition that the O-mode wave damp on the target population of electrons [Eq.~\ref{eq:resE}] at a fixed distance from the onset location of ECRH [Eq.~\ref{eq:onset}].

\subsection{Launcher optimization with HARE}

HARE was originally formulated as a 0-D model for estimating maximum ECCD efficiency given global plasma parameters (on-axis density, temperature, magnetic field, etc.). Here we describe how to extend HARE to be a 1-D model for predicting the maximum ECCD efficiency as a function of the normalized radial variable $\rho$. (The precise definition of $\rho$ is not important for this section.) Clearly, the original HARE model can be directly applied to the magnetic axis $\rho = 0$ without any modifications by simply setting $R_{dep} = R_0$, with $R_0$ the major radius of the magnetic axis, and using as $T_e$, $\omega_p$, and $\omega_c$ the local values at $\rho = 0$.

The case $\rho > 0$ requires more careful consideration, mainly due to the multivaluedness of the mapping $R(\rho)$. While $T_e$ and $\omega_p$ are purely functions of $\rho$ (i.e. they are flux functions), the magnetic field $B$ is not; one must therefore decide where on a given flux surface to apply the HARE formulas. Here we choose to evaluate HARE along the high-field side (HFS) midplane of the tokamak flux surfaces. Mathematically, this means that a given flux value $\rho_j$ maps to the value $R_j$ as
\begin{equation}
    \rho_j \mapsto R_j = \text{min}\left\{ R \, | \, \rho(R, Z_0) = \rho_j \right\}
    ,
\end{equation}

\noindent where $\rho(R, Z)$ is the normalized radial coordinate at $(R, Z)$, and $Z_0$ is the $Z$-value of the magnetic axis. 
This choice is done because it is simple to implement for all flux surfaces, and it is also expected to minimize the impact of trapped particles. The general damping geometry that results from applying the HARE model in this manner, namely, damping on the HFS of the magnetic axis but on the LFS of the cyclotron resonance, is reminiscent of the top-launch experiments performed on the DIII-D tokamak to achieve efficient ECCD~\cite{Chen19,Chen22}. 

The other main modification to the original HARE model that we perform here is to forego the use of a standalone calculation of the ECCD efficiency and instead integrate the HARE predictions within a commercial ray-tracing code, here the GENRAY code~\cite{Smirnov03}. This is primarily born out of practical considerations, as a standalone code to compute the ECCD efficiency based on the Lin--Liu formula~\cite{LinLiu03} is often more difficult to obtain than a commerical ray-tracing code in which Lin--Liu ECCD calculation is typically a standard feature. However, this approach also has the practical benefit of yielding launcher parameters (location and angles) directly as an output.

\begin{figure}
    \centering
    \begin{overpic}[width=0.9\linewidth, trim = {2mm 2mm 2mm 2mm}, clip]{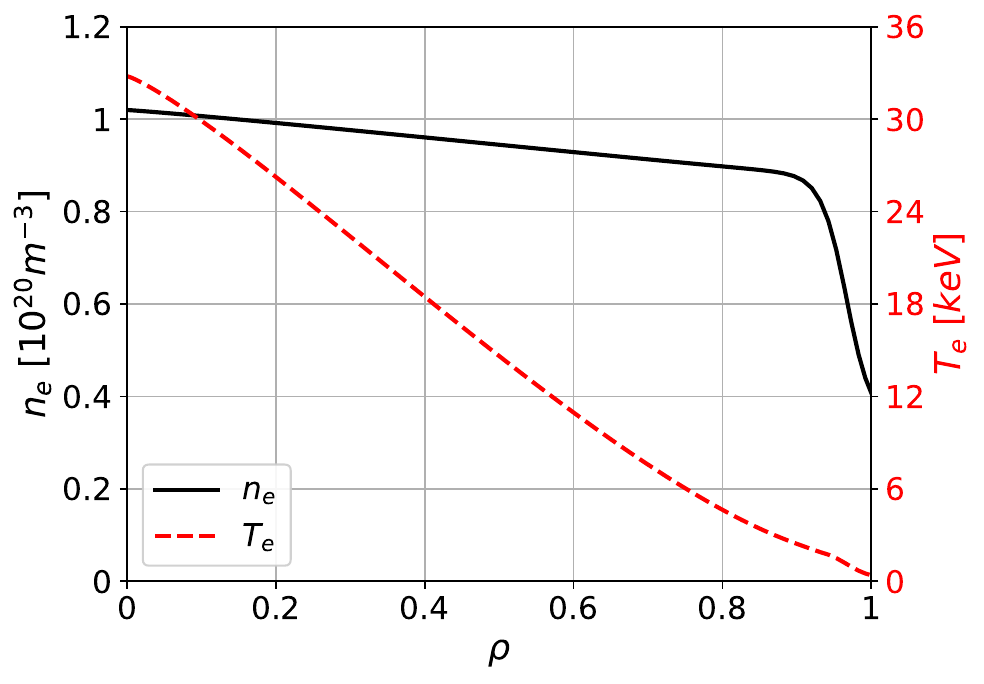}
        \put(1,10){\textbf{(a)}}
    \end{overpic}
    \begin{overpic}[width=0.9\linewidth, trim = {2mm 2mm 2mm 2mm}, clip]{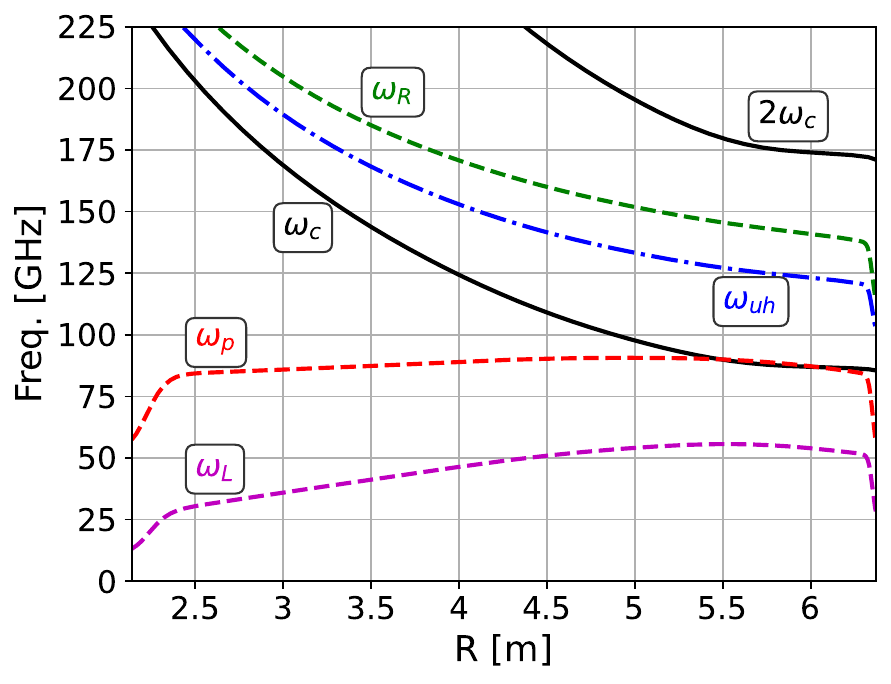}
        \put(1,10){\textbf{(b)}}
    \end{overpic}
    \caption{\textbf{(a)} Electron density $n_e$ and electron temperature $T_e$ profiles as functions of the normalized minor radius $\rho$, with $\rho$ defined in Eq.~\ref{eq:rhoDEF}. \textbf{(b)} Characteristic cutoff and resonance frequencies [see Eqs.~\ref{eq:freqBEG} -- \ref{eq:freqEND}] as functions of major radius along the plasma midplane.}
    \label{fig:profs_freqs_A2}
\end{figure}
 
Specifically, the local HARE predictions are used to initialize an emission process in GENRAY as follows. One first uses the HARE-predicted values of $\omega$ and $N_\parallel$ at a given flux surface (along the HFS midplane, as discussed above) to initialize a ray-tracing simulation from that location, with the ray directed out of the plasma. The initial conditions are, of course, underspecified - fixing the frequency, location and $N_\parallel$, while solving for $|N_\perp|$ via the local dispersion relation still leaves the angle of $N_\perp$ within the plane perpendicular to $B$ undetermined. We choose to remove this degeneracy by imposing that the radial component of $N_\perp$ be zero, i.~e., $N_\rho = 0$. This choice is done to have the rays be tangent to the desired flux surface, thereby maximizing the deposition length of the ray contained within that flux surface. There still remain two possible solutions to these initial conditions, corresponding to whether the ray travels towards the top or bottom half of the vacuum vessel, and we consider both possibilities to accommodate the inherent up-down asymmetry in EC propagation due to the poloidal magnetic field. 

The corresponding launcher configurations are then reconstructed by imposing that the optimal ECCD ray be the time-reversed image of the emitted ray.
A second set of ray-tracing simulations launching rays back into the plasma can then be performed to confirm high ECCD efficiency and localization at the desired deposition location. In practice, only a single ray (rather than a ray bundle) is used in both ray-tracing steps of this optimization workflow for performance considerations, but this choice can also be justified on a conceptual level because most ray-tracing codes, including GENRAY, do not include the complete multi-ray physics anyways~\cite{Lopez20,Lopez21a,Lopez22}. 

\section{Comparison with traditional ECCD launcher optimization}
\label{sec:main}

\begin{table}
\lineup
\caption{HARE-predicted optimal ECCD parameters for the example profiles shown in Fig.~\ref{fig:profs_freqs_A2}.}
\begin{indented}
\item[]\begin{tabular}{ @{} l r r r r r r r}
\br
$\rho$ & $0.0$ & $0.1$ & $0.2$ & $0.3$ & $0.4$ & $0.5$ & $0.6$\\
\mr
$\omega/\omega_c$ & 1.41 & 1.37 & 1.33 & 1.29 & 1.24 & 1.20 & 1.15\\
$|N_\parallel|$ & 0.73 & 0.71 & 0.69 & 0.66 & 0.63 & 0.60 & 0.56\\
$f$~[GHz] & 140 & 154 & 160 & 164 & 169 & 174 & 180\\
\br
\end{tabular}
\end{indented}
\label{tab:HAREfull_A2}
\end{table}

\begin{table}
\lineup
\caption{Same as Table~\ref{tab:HAREfull_A2} but for the simplified HARE model with $|N_\parallel| = 0.7$. The parameters do not differ significantly from full HARE prediction (Table~\ref{tab:HAREfull_A2}) for $\rho \lesssim 0.4$.}
\begin{indented}
\item[]\begin{tabular}{ @{} l r r r r r r r}
\br
$\rho$ & $0.0$ & $0.1$ & $0.2$ & $0.3$ & $0.4$ & $0.5$ & $0.6$\\
\mr
$\omega/\omega_c$ & 1.37 & 1.36 & 1.34 & 1.32 & 1.29 & 1.26 & 1.22\\
$f$~[GHz] & 136 & 153 & 161 & 168 & 175 & 183 & 190\\
\br
\end{tabular}
\end{indented}
\label{tab:HAREsimple_A2}
\end{table}

\begin{figure*}
    \centering
    \begin{overpic}[height=0.5\linewidth, trim = {2mm 2mm 2mm 2mm}, clip]{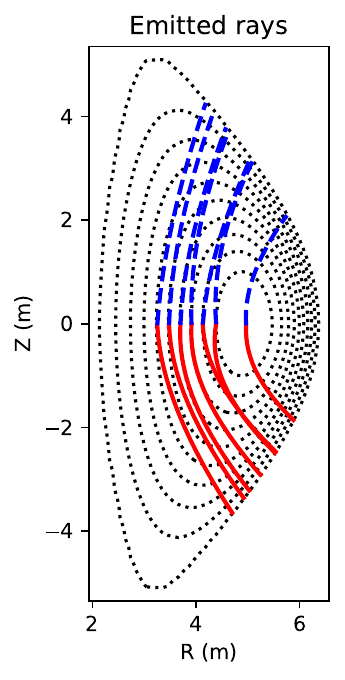}
        \put(40,14){\textbf{(a)}}
    \end{overpic}
    \hspace{2mm}\begin{overpic}[height=0.5\linewidth, trim = {2mm 2mm 2mm 2mm}, clip]{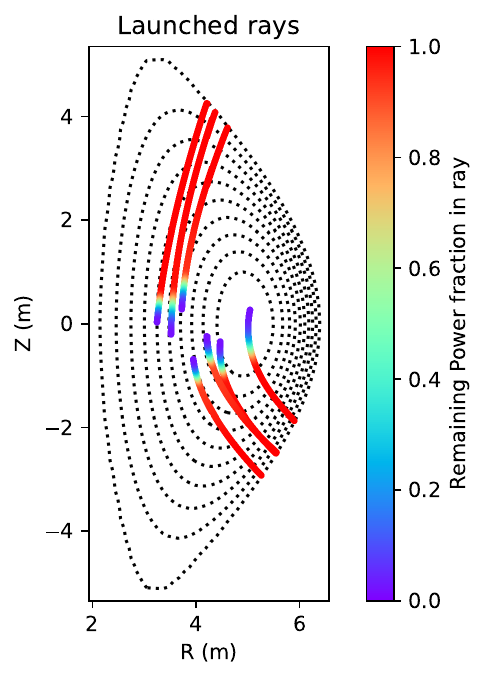}
        \put(40,14){\textbf{(b)}}
    \end{overpic}
    \hspace{2mm}\begin{overpic}[height=0.5\linewidth, trim = {2mm 2mm 2mm 2mm}, clip]{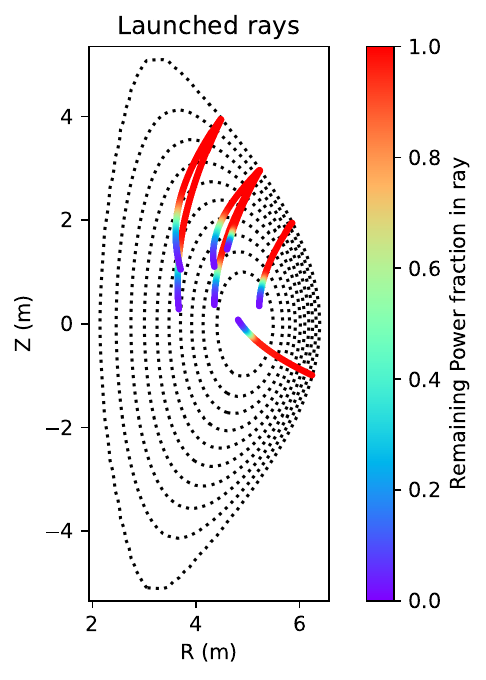}
        \put(40,14){\textbf{(c)}}
    \end{overpic}
    \caption{\textbf{(a)} Emitted rays launched from the desired deposition location with the HARE-predicted optimal parameters (Table~\ref{tab:HAREfull_A2}). The blue dashed lines and the red solid lines represent rays emitted towards the top and bottom of the plasma periphery, respectively. \textbf{(b)} Re-launched rays giving near-optimal ECCD using the exit trajectories determined from the emitted rays, as listed in Table~\ref{tab:ECCDparams_HAREfull_A2}. \textbf{(c)} Nearly-optimal ECCD ray trajectories obtained using the traditional approach based on coarse parameter scan of initial launcher conditions, as listed in Table~\ref{tab:ECCDparams_traditional_A2}.}
    \label{fig:rays_A2}
\end{figure*}

\begin{table*}
\lineup
\caption{Optimal ECCD launcher obtained via reverse ray-tracing for standard HARE parameters (Table~\ref{tab:HAREfull_A2}) on the example profiles shown in Fig.~\ref{fig:profs_freqs_A2}. Note that $R_{st}$ and $Z_{st}$ are the radial and vertical position of the launcher, respectively, $\alpha_{st}$ and $\beta_{st}$ are the toroidal and poloidal launch angles, respectively, and $\Delta \rho$ is the difference between the obtained and desired peak deposition locations.}
\begin{indented}
\item[]\begin{tabular}{ @{} l r r r r r r r}
\br
$\rho$ & $0.0$ & $0.1$ & $0.2$ & $0.3$ & $0.4$ & $0.5$ & $0.6$\\
\mr
$f$~[GHz] & 140 & 154 & 160 & 164 & 169 & 174 & 180\\
$R_{st}$~[cm] & 588 & 554 & 555 & 526 & 460 & 436 & 421\\
$Z_{st}$~[cm] & -186 & -250 & -248 & -293 & 380 & 411 & 428\\
$\alpha_{st}$~[$^\circ$] & -48 & -47 & -44 & -44 & -49 & -49 & -45\\
$\beta_{st}$~[$^\circ$] & 32 & 33 & 33 & 36 & -53 & -57 & -60\\
$\Delta \rho$ & 0.003 & 0.023 & -0.023 & 0.018 & -0.003 & -0.018 & -0.008 \\
$\zeta$ & 0.341 & 0.316 & 0.291 & 0.296 & 0.309 & 0.339 & 0.341\\
\mr
$I/P$~[kA/MW] & 78.1 & 65.4 & 56.5 & 47.0 & 42.6 & 39.0 & 29.3\\
\br
\end{tabular}
\end{indented}
\label{tab:ECCDparams_HAREfull_A2}
\end{table*}

\begin{figure}
    \centering
    \includegraphics[width=\linewidth, trim = {2mm 2mm 2mm 2mm}, clip]{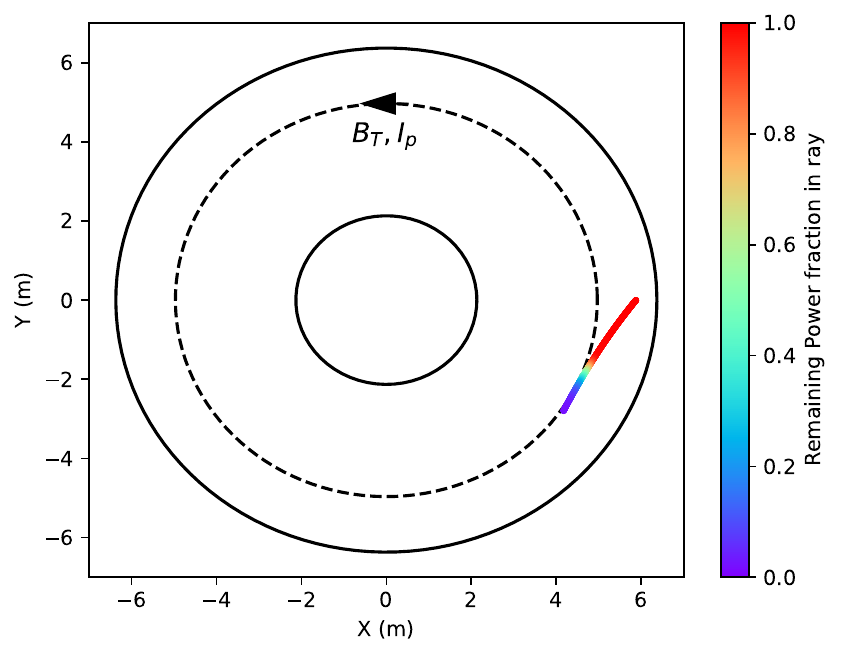}
    \caption{Toroidal projection of the launched EC ray trajectory. Notably, both the toroidal magnetic field $B_T$ and the plasma current $I_p$ are counter-clockwise when viewed from above, so $N_\parallel < 0$. This feature is common for all cases considered here.}
    \label{fig:toroidal}
\end{figure}

In this section we apply the new physics-based ECCD optimization scheme described in the previous section to a series of two example tokamak equilibria with reactor-relevant parameters%
\footnote{Since our goal is to compare two different EC launcher optimization methods, the precise origin of the plasma profiles is relatively unimportant; that said, the profiles we consider have been obtained from integrated modeling of two different design points under consideration for a possible ST-based FPP, whose details will be described elsewhere in a dedicated publication.}%
, and compare the results to the traditional method of optimizing ECCD launchers based on a coarse sampling of the possible launcher initial conditions (frequency, location, and launch angles). Since the second example ultimately will show qualitatively similar performance results as the first example, it is discussed more briefly to avoid belaboring the main point; the hurried reader can also simply skip the second example without too much narrative loss.

\subsection{Example 1: Aspect ratio A = 2 equilibrium}

The first example profile is shown in Fig.~\ref{fig:profs_freqs_A2}. As shown, this equilibrium has central values $T_e \approx 33$~keV and $n_e \approx 1.02 \times 10^{20}$~$\text{m}^{-3}$. The flux coordinate used is the normalized poloidal flux:
\begin{equation}
    \rho
    =
    \frac{
        \Psi - \Psi_0
    }{
        \Psi_{LCFS} - \Psi_0
    }
    ,
    \label{eq:rhoDEF}
\end{equation}

\noindent where $\Psi$ is the poloidal magnetic flux, $\Psi_0$ is the poloidal magnetic flux at the magnetic axis, and $\Psi_{LCFS}$ is the poloidal magnetic flux at the LCFS. Also shown are the fundamental cyclotron resonance frequency $\omega_c$, the plasma cutoff frequency $\omega_p$, the upper-hybrid resonance frequency $\omega_{uh}$, the right-hand cutoff frequency $\omega_R$, and the left-hand cutoff frequency $\omega_L$, defined respectively as
\numparts
    \begin{eqnarray}
        \label{eq:freqBEG}
        \omega_c &= 28 \left( \frac{B}{1~\text{Tesla}} \right)~\text{GHz}
        , \\
        \label{eq:freqPLASMA}
        \omega_p &= 89.8 \sqrt{ \frac{n_e}{10^{20}~\text{m}^{-3}} }~\text{GHz}
        , \\
        \omega_{uh} &= \sqrt{ \omega_c^2 + \omega_p^2}
        , \\
        \omega_R &= 
        \sqrt{ \omega_p^2 + \frac{\omega_c^2}{4} }
        + \frac{\omega_c}{2}
        , \\
        \label{eq:freqEND}
        \omega_L &= 
        \sqrt{ \omega_p^2 + \frac{\omega_c^2}{4} }
        - \frac{\omega_c}{2}
        ,
    \end{eqnarray}
\endnumparts

\noindent where, as indicated, the magnetic field should be evaluated in Tesla and the electron density should be evaluated in units of $10^{20}~\text{m}^{-3}$.

As shown in the figure, fundamental O-mode heating is accessible for nearly all frequencies present in the plasma, whereas the second-harmonic X-mode heating requires impractically high frequencies to cover the entire plasma volume. Therefore, we can restrict attention to fundamental O-mode heating, as assumed in the HARE model presented in Sec.~\ref{sec:background}.

\begin{table}
\lineup
\caption{Same as Table~\ref{tab:ECCDparams_HAREfull_A2} but for the simplified HARE model (Table~\ref{tab:HAREsimple_A2}). Note that $\rho < 0.4$ is not considered, as the parameters will not change significantly from those presented in Table~\ref{tab:ECCDparams_HAREfull_A2}.}
\begin{indented}
\item[]\begin{tabular}{ @{} l r r r }
\br
$\rho$
& $0.4$ & $0.5$ & $0.6$\\
\mr
$f$~[GHz] & 175 & 183 & 190\\
$R_{st}$~[cm]
& 488 & 478 & 467 \\
$Z_{st}$~[cm]
& 346 & 359 & 373 \\
$\alpha_{st}$~[$^\circ$] 
& -46 & -43 & -41 \\
$\beta_{st}$~[$^\circ$]
& -48 & -49 & -50 \\
$\Delta \rho$
& -0.028 & -0.013 & -0.023 \\
$\zeta$
& 0.311 & 0.317 & 0.335 \\
\mr
$I/P$~[kA/MW]
& 44.9 & 36.0 & 30.1 \\
\br
\end{tabular}
\end{indented}
\label{tab:ECCDparams_HAREsimple_A2}
\end{table}

The optimal launching frequency $f = \omega/2\pi$ and parallel refractive index $N_\parallel$ at the desired damping location $\rho$ as predicted by the $1$-D HARE model (Eqs.~\ref{eq:HAREfullOMEGA} and \ref{eq:HAREfullNpar}, but see also Sec.~\ref{sec:HARE} for the complete calculation workflow) are shown in Table~\ref{tab:HAREfull_A2}; while the optimal $f$ as predicted by the simplified HARE model assuming $|N_\parallel| = 0.7$ at the damping location (Eq.~\ref{eq:HAREsimpleOMEGA}) are shown in Table~\ref{tab:HAREsimple_A2}. Only radial locations relatively close to the core of the plasma equilibrium ($\rho \le 0.6)$ are considered because beyond this range, the rise in bootstrap current will lessen the need for ECCD. We also choose to sample $\rho$ with stepsize $0.1$ for simplicity. 

It is seen that within this range of deposition locations, the complete and the simplified HARE models only differ significantly for $\rho \gtrsim 0.4$. In this outer region, the simplified HARE model returns a higher frequency than the complete HARE model because the imposed $|N_\parallel| = 0.7$ of the former is larger than the $N_\parallel$ calculated by the latter; higher $N_\parallel$ then implies a larger Doppler-shifted frequency to compensate.

\begin{figure}
    \centering
    \begin{overpic}[width=0.9\linewidth, trim = {2mm 2mm 2mm 2mm}, clip]{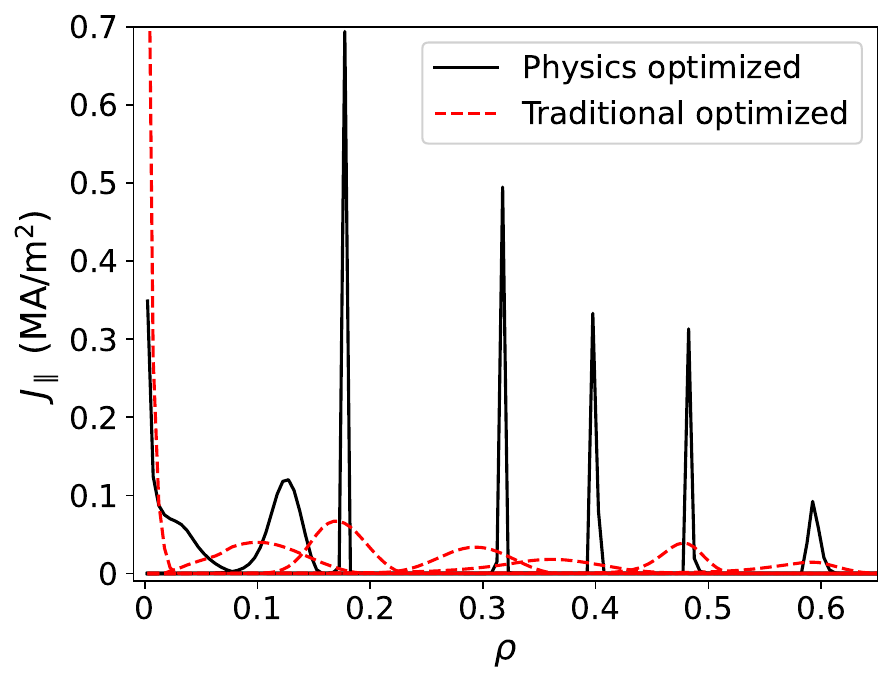}
        \put(1,10){\textbf{(a)}}
    \end{overpic}

    \begin{overpic}[width=0.9\linewidth, trim = {2mm 2mm 2mm 2mm}, clip]{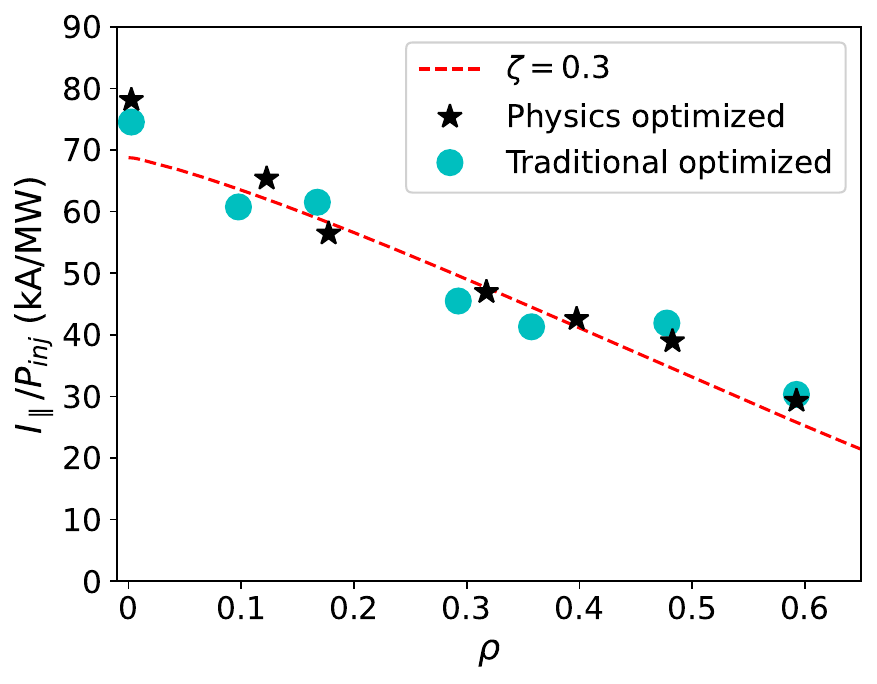}
        \put(1,10){\textbf{(b)}}
    \end{overpic}
    \caption{\textbf{(a)} Comparison of the obtained ECCD profiles using the new physics-based optimization scheme and the traditional optimization scheme based on coarse parameter scan. \textbf{(b)} Comparison of the ECCD efficiency (total driven current per total injected power) for the two optimization methods. Also shown is the corresponding curve [Eq.~\ref{eq:ECCDtotal}] for when the non-dimensional ECCD efficiency $\zeta = 0.3$, with $\zeta$ defined in Eq.~\ref{eq:ECCDeffic}.}
    \label{fig:ECCD_A2}
\end{figure}

As shown in Fig.~\ref{fig:rays_A2}, the optimal HARE parameters are used to initialize a pair of ray-tracing simulations emitting rays towards the top and bottom half of the plasma periphery. The optimal launching parameters are then obtained from the exit parameters of these emitted rays, and used to initialize a second set of ray-tracing simulations to re-launch the EC rays back into the plasma and obtain the estimated optimal ECCD efficiency. After comparing the performance of the top and bottom-half launchers, the optimal set of launching parameters obtained by this method for both the complete HARE model and the simplified HARE model are shown in Tables~\ref{tab:ECCDparams_HAREfull_A2} and \ref{tab:ECCDparams_HAREsimple_A2}, respectively. Note that for all cases considered here, the toroidal magnetic field and the plasma current are both oriented counter-clockwise when viewed from above, so driving positive current requires the EC launchers to have $N_\parallel < 0$; this is shown in Fig.~\ref{fig:toroidal}.

The EC launch parameters obtained by both HARE models result in comparable ECCD performance. Both models achieve a normalized dimensionless ECCD efficiency of $\zeta \approx 0.3$ throughout the considered plasma volume, where $\zeta$ is defined as~\cite{Luce99a}
\numparts
    \begin{eqnarray}
        \label{eq:ECCDeffic}
        \zeta
        \approx 
        0.033
        \left( \frac{I_p/ P_{in}}{1~\text{kA/MW}} \right)
        \left( \frac{R_{geo} n_e}{10^{20}~\text{m}^{-2}} \right) 
        \left( \frac{1~\text{keV} }{T_e} \right)
        , \\
        \label{eq:ECCDtotal}
        \frac{I_p}{P_{in}}
        \approx
        30.3
        \, \zeta
        \left( \frac{T_e}{1~\text{keV} } \right)
        \left( \frac{10^{20}~\text{m}^{-2}}{R_{geo} n_e} \right)
        ~\text{kA/MW}
        ,
    \end{eqnarray}
\endnumparts

\begin{table*}
\lineup
\caption{Same as Table~\ref{tab:ECCDparams_HAREfull_A2} but for a coarse parameter scan of $(f, R_{st}, Z_{st}, \alpha_{st}, \beta_{st})$ parameter space.}
\begin{indented}
\item[]\begin{tabular}{ @{} l r r r r r r r}
\br
$\rho$ & $0.0$ & $0.1$ & $0.2$ & $0.3$ & $0.4$ & $0.5$ & $0.6$\\
\mr
$f$~[GHz] & 140 & 130 & 160 & 140 & 160 & 180 & 185\\
$R_{st}$~[cm] & 625 & 590 & 525 & 525 & 525 & 450 & 450\\
$Z_{st}$~[cm] & -100 & 200 & 300 & 300 & 300 & 400 & 400\\
$\alpha_{st}$~[$^\circ$] & -40 & -50 & -50 & -50 & -50 & -50 & -50\\
$\beta_{st}$~[$^\circ$] & 20 & -30 & -40 & -40 & -30 & -49 & -40\\
$\Delta \rho$ & 0.003 & -0.003 & -0.033 & -0.008 & -0.043 & -0.023 & -0.008 \\
$\zeta$ & 0.325 & 0.286 & 0.313 & 0.275 & 0.278 & 0.360 & 0.353\\
\mr
$I/P$~[kA/MW] & 74.6 & 60.8 & 61.5 & 45.5 & 41.3 & 41.9 & 30.3\\
\br
\end{tabular}
\end{indented}
\label{tab:ECCDparams_traditional_A2}
\end{table*}

\noindent with $I_p$ the total driven EC current and $P_{in}$ the total injected EC power. Both models also exhibit the same trend that bottom launchers are favorable for near-central ECCD ($\rho \lesssim 0.3$), while top launchers are preferred for more peripheral damping locations. This is largely attributed to the evolution of $N_\parallel$ along the rays - for both this example plasma and the following example plasma, the plasma current is co-directional with respect to the toroidal magnetic field such that the magnetic field lines circulate clockwise in the poloidal plane; it is therefore easier to maintain a nearly constant (negative) value of $N_\parallel$ when launching from the lower half (resp.~upper half) to deposit near the center (resp.~periphery along HFS midplane).

The fact that the simplified HARE model gives comparable ECCD with the complete HARE model suggests that achieving efficient ECCD is not extremely sensitive to the launching parameters; the operational window can be quickly estimated by comparing the two model outputs, as in Tables~\ref{tab:ECCDparams_HAREfull_A2} and \ref{tab:ECCDparams_HAREsimple_A2}. Additionally, this means that choosing the underlying HARE model depends largely on additional considerations besides ECCD performance. For example, the complete HARE model allows more flexibility over the localization of the ECCD profile through the stronger dependence on $\Delta$. However, the simplified HARE model might be more amenable to engineering constraints because it predicts less quantities; one could imagine fixing the launch frequency to a specified value and solving for the optimal $N_\parallel$ via Eq.~\ref{eq:HAREsimpleOMEGA} instead of prescribing $N_\parallel$ as we do here.

For this example plasma profile, we also obtained estimates for the optimal ECCD launcher parameters via the traditional approach. This involves a large, albeit coarse, parameter scan in which the wave frequency was scanned over the range $[80~\text{GHz}, 220~\text{GHz}]$ at $5$~GHz increments, the toroidal and poloidal injection angles were both scanned over the range $[-90^\circ, 90^\circ]$ at $10^\circ$ increments, and the launcher location was varied between $9$ candidate positions. All together, approximately one hundred thousand GENRAY simulations were required to complete this parameter scan. The resulting estimates for the optimal launcher configurations to drive ECCD at each radial location is shown in Table~\ref{tab:ECCDparams_traditional_A2}. In particular, it is seen that launchers achieving $\zeta \approx 0.3$ throughout the considered plasma volume are also obtained via this traditional approach.

\begin{figure}
    \centering
    \begin{overpic}[width=0.9\linewidth, trim = {2mm 2mm 2mm 2mm}, clip]{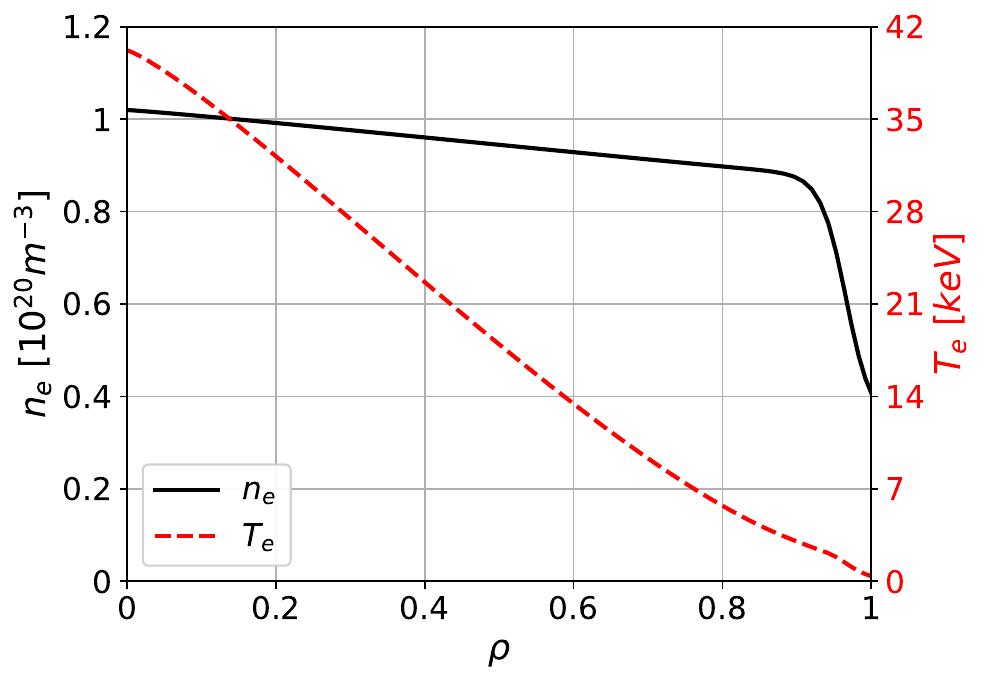}
        \put(1,10){\textbf{(a)}}
    \end{overpic}
    \begin{overpic}[width=0.9\linewidth, trim = {2mm 2mm 2mm 2mm}, clip]{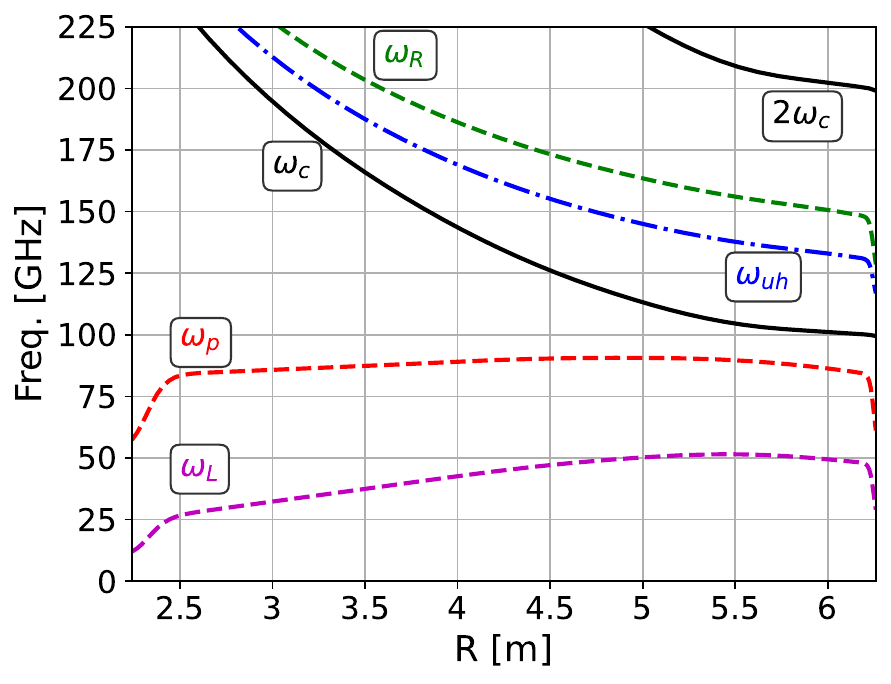}
        \put(1,10){\textbf{(b)}}
    \end{overpic}
    \caption{Same as Fig.~\ref{fig:profs_freqs_A2}, but for the second example profiles.}
    \label{fig:profs_freqs_A21}
\end{figure}

The ECCD profiles and total driven current as functions of $\rho$ obtained from the new physics-based optimization method and the traditional optimization method are shown in Fig.~\ref{fig:ECCD_A2}. Both methods can drive a similar amount of total current across the plasma radius, achieving $\zeta \approx 0.3$ with comparable amount of scatter. This is a particularly promising result because the new method achieves this using only $28$ simulations (2 outgoing + 2 incoming simulations to obtain top and bottom launcher candidates for each $7$ radial location considered), compared to the aforementioned $100,000$ simulations for the traditional result. The ECCD profiles obtained by the new approach are also generally more localized compared to the traditionally optimized launchers. This is by design according to our specific choices for the free parameters in the HARE model and imposed damping geometry; different choices can likely broaden the ECCD profiles obtained with the new approach, which will be explored in future investigations. Conversely, there is no direct way to impose the ECCD localization in the traditional optimization method - one must simply hope that the coarse parameter scan yields a localized profile at a given desired location, or refine the scan and repeat if not.

\begin{table}
\lineup
\caption{Same as Table~\ref{tab:HAREfull_A2} but for the example profiles shown in Fig.~\ref{fig:profs_freqs_A21}.}
\begin{indented}
\item[]\begin{tabular}{ @{} l r r r r r r r}
\br
$\rho$ & $0.0$ & $0.1$ & $0.2$ & $0.3$ & $0.4$ & $0.5$ & $0.6$\\
\mr
$\omega/\omega_c$ & 1.47 & 1.43 & 1.38 & 1.32 & 1.27 & 1.22 & 1.17\\
$|N_\parallel|$ & 0.75 & 0.73 & 0.71 & 0.69 & 0.65 & 0.62 & 0.58\\
$f$~[GHz] & 169 & 187 & 193 & 196 & 199 & 203 & 209\\
\br
\end{tabular}
\end{indented}
\label{tab:HAREfull_A21}
\end{table}

\begin{table}
\lineup
\caption{Same as Table~\ref{tab:HAREsimple_A2} but for the example profiles shown in Fig.~\ref{fig:profs_freqs_A21}. The parameters do not differ significantly from full HARE prediction (Table~\ref{tab:HAREfull_A21}) for $0.1 \lesssim \rho \lesssim 0.4$.}
\begin{indented}
\item[]\begin{tabular}{ @{} l r r r r r r r}
\br
$\rho$ & $0.0$ & $0.1$ & $0.2$ & $0.3$ & $0.4$ & $0.5$ & $0.6$\\
\mr
$\omega/\omega_c$ & 1.39 & 1.38 & 1.36 & 1.34 & 1.31 & 1.28 & 1.24\\
$f$~[GHz] & 159 & 181 & 190 & 198 & 206 & 213 & 220\\
\br
\end{tabular}
\end{indented}
\label{tab:HAREsimple_A21}
\end{table}

\begin{figure*}
    \centering
    \begin{overpic}[height=0.49\linewidth, trim = {2mm 2mm 2mm 2mm}, clip]{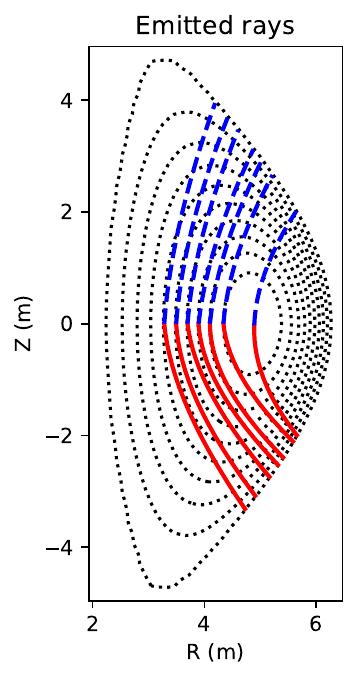}
        \put(40,14){\textbf{(a)}}
    \end{overpic}
    \hspace{2mm}\begin{overpic}[height=0.49\linewidth, trim = {2mm 2mm 2mm 2mm}, clip]{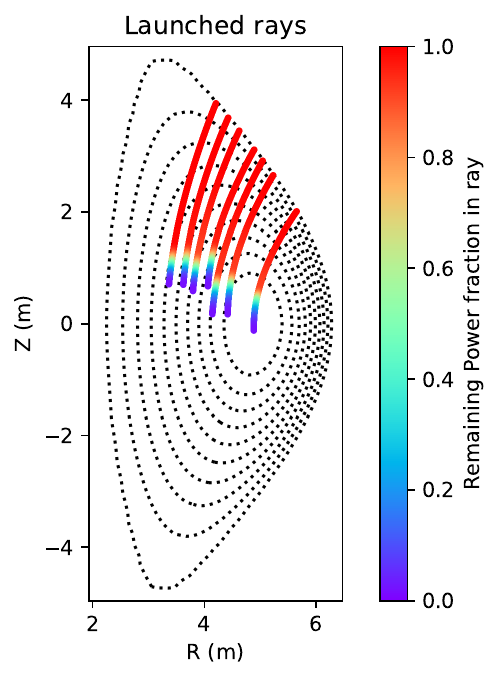}
        \put(40,14){\textbf{(b)}}
    \end{overpic}
    \hspace{2mm}\begin{overpic}[height=0.49\linewidth, trim = {2mm 2mm 2mm 2mm}, clip]{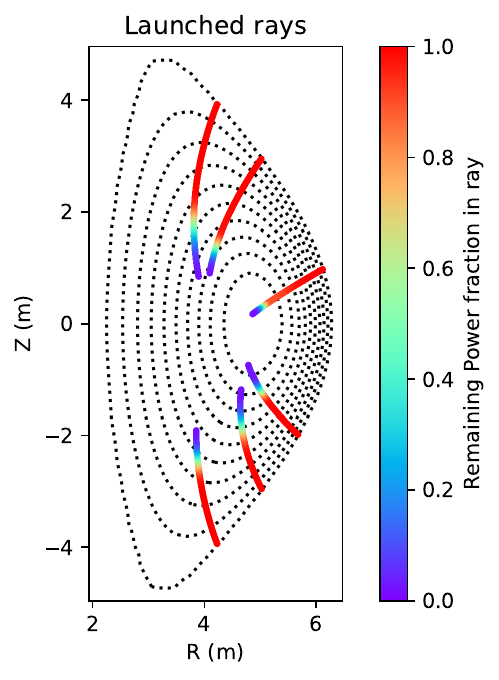}
        \put(40,14){\textbf{(c)}}
    \end{overpic}
    \caption{Same as Fig.~\ref{fig:rays_A2} but for the example profiles shown in Fig.~\ref{fig:profs_freqs_A21}.}
    \label{fig:rays_A21}
\end{figure*}

\begin{table*}
\lineup
\caption{Same as Table~\ref{tab:ECCDparams_HAREfull_A2} but for the example profiles shown in Fig.~\ref{fig:profs_freqs_A21}.}
\begin{indented}
\item[]\begin{tabular}{ @{} l r r r r r r r}
\br
$\rho$ & $0.0$ & $0.1$ & $0.2$ & $0.3$ & $0.4$ & $0.5$ & $0.6$\\
\mr
$f$~[GHz] & 169 & 187 & 193 & 196 & 199 & 203 & 209\\
$R_{st}$~[cm] & 566 & 524 & 505 & 490 & 463 & 443 & 421\\
$Z_{st}$~[cm] & 203 & 268 & 294 & 314 & 348 & 371 & 397\\
$\alpha_{st}$~[$^\circ$] & -53 & -53 & -51 & -52 & -51 & -51 & -50\\
$\beta_{st}$~[$^\circ$] & -35 & -41 & -45 & -44 & -50 & -52 & -56\\
$\Delta \rho$ & 0.003 & 0.008 & -0.008 & 0.003 & -0.003 & -0.013 & -0.008 \\
$\zeta$ & 0.291 & 0.285 & 0.280 & 0.295 & 0.288 & 0.312 & 0.332\\
\mr
$I/P$~[kA/MW] & 81.6 & 73.5 & 65.5 & 59.0 & 48.6 & 43.6 & 35.0\\
\br
\end{tabular}
\end{indented}
\label{tab:ECCDparams_HAREfull_A21}
\end{table*}

\subsection{Example 2: Aspect ratio A = 2.1 equilibrium}

For further validation of the new optimization method, we can repeat the analysis on a second example plasma profile, shown in Fig.~\ref{fig:profs_freqs_A21}. This second example plasma is similar to the first, but notably has a higher magnetic field and also a higher central $T_e \approx 40$~keV. As a result, the optimal EC frequencies are expected to be higher for this second example plasma compared to the first example. This is confirmed by the HARE-predicted optimal parameters listed in Tables~\ref{tab:HAREfull_A21} and \ref{tab:HAREsimple_A21} using the complete and simplified model formulations, respectively. Analogous to the first example, for the second plasma profile the simplified HARE parameters do not vary significantly from the complete HARE parameters when $0.1 \lesssim \rho \lesssim 0.4$.

\begin{table}
\lineup
\caption{Same as Table~\ref{tab:ECCDparams_HAREsimple_A2} but for the example profiles shown in Fig.~\ref{fig:profs_freqs_A21}.}
\begin{indented}
\item[]\begin{tabular}{ @{} l r r r r r }
\br
$\rho$ & $0.0$ & $0.1$ & $0.4$ & $0.5$ & $0.6$\\
\mr
$f$~[GHz] & 159 & 181 & 206 & 213 & 220\\
$R_{st}$~[cm] & 554 & 510 & 475 & 466 & 456 \\
$Z_{st}$~[cm] & 223 & 287 & 332 & 344 & 356 \\
$\alpha_{st}$~[$^\circ$] & -54 & -54 & -49 & -48 & -46 \\
$\beta_{st}$~[$^\circ$] & -40 & -45 & -46 & -45 & -45 \\
$\Delta \rho$ & 0.003 & -0.008 & 0.008 & -0.008 & -0.018 \\
$\zeta$ & 0.255 & 0.254 & 0.302 & 0.318 & 0.328 \\
\mr
$I/P$~[kA/MW] & 71.5 & 66.4 & 50.0 & 43.8 & 35.7 \\
\br
\end{tabular}
\end{indented}
\label{tab:ECCDparams_HAREsimple_A21}
\end{table}

Like in the previous example, the optimal HARE parameters are used to emit rays toward the top and bottom of the plasma periphery, the corresponding launcher parameters are extracted, and the rays are re-launched back into the plasma to obtain the corresponding ECCD profiles and efficiencies. The relevant ray trajectories are shown in Fig.~\ref{fig:rays_A21}, and the obtained launcher parameters and performance for the complete and simplified HARE models are shown in Tables~\ref{tab:ECCDparams_HAREfull_A21} and \ref{tab:ECCDparams_HAREsimple_A21}, respectively.

Similar to the first example profile, the new optimization method yields launchers that can achieve $\zeta \approx 0.3$ across the plasma radius. Unlike the first example, however, for these profiles the top launchers perform significantly better than the bottom launchers. This is because the higher $T_e$ means there is increased parasitic absorption at the second harmonic resonance; the smaller radial location of the top launchers compared to the bottom launchers (see Fig.~\ref{fig:rays_A21}) means the top launchers are less susceptible to this effect.

\begin{table*}
\lineup
\caption{Same as Table~\ref{tab:ECCDparams_traditional_A2} but for the example profiles shown in Fig.~\ref{fig:profs_freqs_A21}.}
\begin{indented}
\item[]\begin{tabular}{ @{} l r r r r r r r}
\br
$\rho$ & $0.0$ & $0.1$ & $0.2$ & $0.3$ & $0.4$ & $0.5$ & $0.6$\\
\mr
$f$~[GHz] & 160 & 170 & 165 & 195 & 170 & 200 & 180\\
$R_{st}$~[cm] & 615 & 570 & 570 & 505 & 505 & 425 & 425\\
$Z_{st}$~[cm] & 100 & -200 & -200 & 300 & -300 & 400 & -400\\
$\alpha_{st}$~[$^\circ$] & -40 & -50 & -50 & -50 & -60 & -60 & -60\\
$\beta_{st}$~[$^\circ$] & -20 & 30 & 30 & -40 & 40 & -49 & 50\\
$\Delta \rho$ & 0.008 & 0.028 & -0.023 & 0.018 & -0.023 & 0.023 & 0.023 \\
$\zeta$ & 0.250 & 0.282 & 0.270 & 0.288 & 0.294 & 0.352 & 0.314\\
\mr
$I/P$~[kA/MW] & 70.2 & 71.0 & 64.2 & 56.2 & 51.5 & 45.1 & 30.2 \\
\br
\end{tabular}
\end{indented}
\label{tab:ECCDparams_traditional_A21}
\end{table*}

\begin{figure}
    \centering
    \begin{overpic}[width=0.9\linewidth, trim = {2mm 2mm 2mm 2mm}, clip]{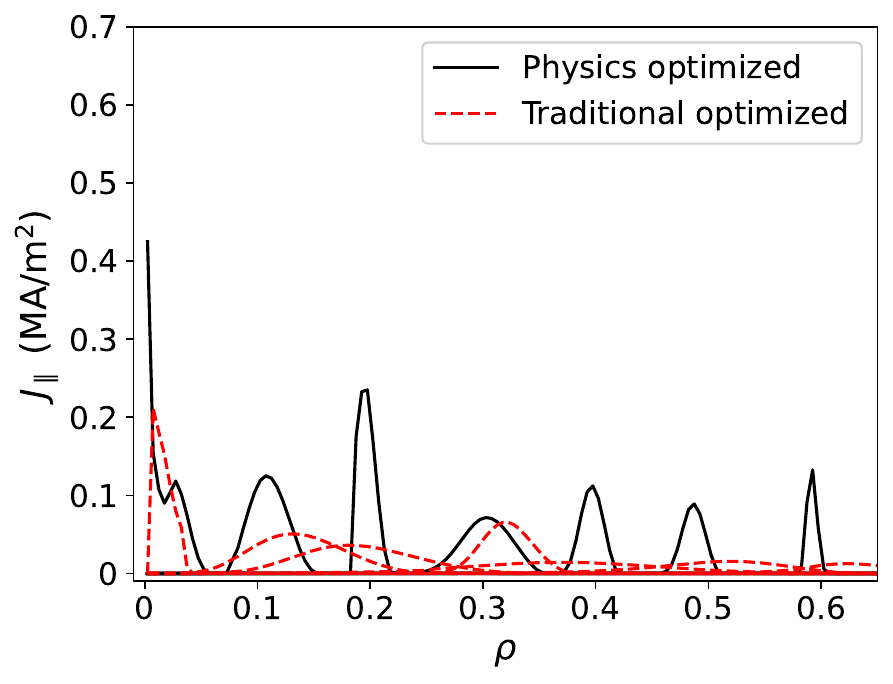}
        \put(1,10){\textbf{(a)}}
    \end{overpic}

    \begin{overpic}[width=0.9\linewidth, trim = {2mm 2mm 2mm 2mm}, clip]{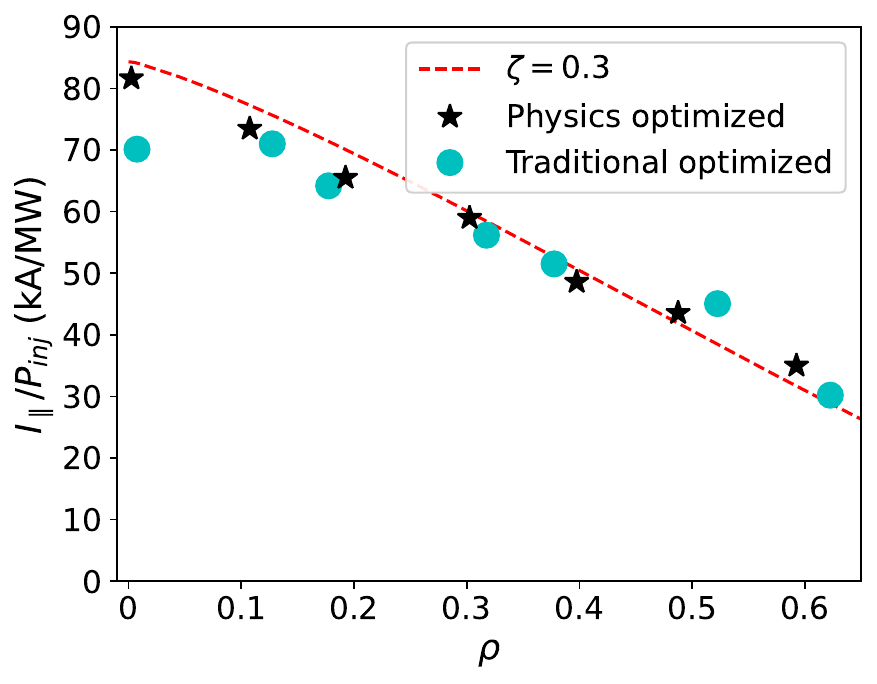}
        \put(1,10){\textbf{(b)}}
    \end{overpic}
    \caption{Same as Fig.~\ref{fig:ECCD_A2} but for the example profiles shown in Fig.~\ref{fig:profs_freqs_A21}.}
    \label{fig:ECCD_A21}
\end{figure}

The same coarse parameter scan is performed on this example plasma profile as was done on the previous example; the resulting estimates for the optimal launcher parameters at each radial location are shown in Table~\ref{tab:ECCDparams_traditional_A21}. Again, the coarse parameter scan reproduces the main finding of the physics-based method, namely that there exists a launcher configuration to achieve efficient ($\zeta \approx 0.3$) ECCD at any given flux surface, but obtaining this conclusion via the traditional approach requires approximately $10,000\times$ more simulations. The resulting ECCD profiles obtained by the two optimization methods are further compared in Fig.~\ref{fig:ECCD_A21}. Once again, the new optimization method yields launchers that drive more localized current compared to the traditional approach. The scatter about the optimal ECCD efficiency $\zeta \approx 0.3$ is also comparable between the two methods, further validating the new approach as a faster means of optimizing ECCD launchers.

\section{Conclusion}
\label{sec:concl}

In this work we present a physics-based method for performing initial `zeroth-order' optimization of ECCD launchers for use in tokamak design studies. By `zeroth-order' we refer to the first stage in the design process in which only rough estimates of launcher parameters and the ECCD efficiency profiles are needed; as the designs get more refined, these estimates can be used to initialize more thorough optimizations that begin to account for engineering constraints such as launcher port locations, available frequencies, etc. Traditionally, this `zeroth-order' ECCD launcher optimization is performed via coarsely sampling the possible space of launcher positions, frequencies, and launch angles; more refined parameter scans are then performed locally around the maxima returned by the coarse scans if desired. This method requires many simulations to adequately cover the plasma volume, possibly on the order of millions or more, whereas our approach presented here requires only about one hundred simulations to achieve comparable performance. This is demonstrated on two example plasma profiles with reactor-relevant parameters. Both methods show that normalized ECCD efficiency $\zeta \approx 0.3$ can be achieved across the plasma radius, with comparable scatter arising from the fact that both methods are intended only as initial `zeroth-order' estimates of the optimal parameters. 

The framework we present here improves upon the original underlying HARE model~\cite{EPoli18b} by (1) upgrading the HARE model to 1-D instead of 0-D to allow one to specify the deposition location of interest; (2) not requiring one to have a standalone Lin--Liu ECCD module but instead using commercial ray-tracing codes, which are more practically obtainable; and (3) outputing the launcher parameters needed for optimal ECCD. Additionally, the new physics-based method is capable of driving very localized EC current; as such, it may be useful more generally as a controller framework for rampup design and neoclassical tearing mode control by providing a reliable mapping between deposition location, local plasma parameters, and external launch conditions. Modifications to our methodology, such as changing the free parameters within the HARE model or considering different damping geometries than the high-field-side midplane damping that we consider here, might be used in more sophisticated ECCD optimizations that are not solely focused on efficiency but also account for additional parameters such as ECCD profile width.

To conclude, it is worthwhile to place the ten-thousandfold speedup we achieve here in practical terms. With the standard settings employed here, each GENRAY simulation takes about $10$~seconds; using our approach versus the traditional approach amounts to obtaining optimal launcher estimates in a matter of minutes ($\sim 17$~minutes for 100 simulations) versus a few ($\sim 3.8$) months of CPU time. (Note that these estimates are all for a single ray, i.e., time per ray; using multiple rays will subsequently introduce an overall multiplicative factor.) Said another way, a ten-thousandfold speedup means that more comprehensive physics models, such as adjoint method for ECCD~\cite{Antonsen82,Fisch87} with fully relativistic propagation and absorption~\cite{Ram05,NelsonMelby07} and Westerhof--Tokman procedure~\cite{Westerhof97,Tokman00}, can be incorporated into the initial optimization for the same amount of CPU time, especially if recently developed fast algorithms~\cite{Biswas25} are also used.

\section*{Data availability statement}
The data that supports the findings of this study are available upon reasonable request from the authors.

\ack
This work was prepared as an account of work sponsored by an agency of the United States Government. Neither the United States Government nor any agency thereof, nor any of their employees, nor any of their contractors, subcontractors or their employees, makes any warranty, express or implied, or assumes any legal liability or responsibility for the accuracy, completeness, or any third party’s use or the results of such use of any information, apparatus, product, or process disclosed, or represents that its use would not infringe privately owned rights. Reference herein to any specific commercial product, process, or service by trade name, trademark, manufacturer, or otherwise, does not necessarily constitute or imply its endorsement, recommendation, or favoring by the United States Government or any agency thereof or its contractors or subcontractors. The views and opinions of authors expressed herein do not necessarily state or reflect those of the United States Government or any agency thereof, its contractors or subcontractors.

\section*{References}
\bibliography{Biblio.bib}
\bibliographystyle{iopart-num}

\end{document}